\documentstyle[aps,prl,preprint,floats,epsfig]{revtex}

\begin{document}

\tighten

\preprint{\tighten\vbox{\hbox{\bf CLNS 00/1666 }
			\hbox{\bf CLEO 00-4}
                        \hbox{\date{\today}}}}

\title{\boldmath 
Resonance structure of $\tau^-\to K^-\pi^+\pi^-\nu_\tau$ decays}

\author{CLEO Collaboration}
\date{\today}

\maketitle
\tighten

\begin{abstract} 
Using a sample of 4.7 fb$^{-1}$ integrated luminosity accumulated
with the CLEO~II detector at the Cornell Electron Storage Ring (CESR),
we investigate the mass spectrum and resonant structure in  
$\tau^-\to K^-\pi^+\pi^-\nu_\tau$ decays.
We measure the relative
fractions of $K_1(1270)$ and $K_1(1400)$ resonances in these
decays, as well as the $K_1$ masses and widths.
Our fitted $K_1$ resonances
are somewhat broader than previous hadroproduction measurements,
and in agreement with recent LEP results from tau decay.
The larger central value of our measured width supports models which
attribute the small
$\tau^-\to K^-\pi^+\pi^-\nu_\tau$ branching fraction 
to larger $K_1$ widths than are presently tabulated.
We also determine 
the $K_a-K_b$ mixing angle $\theta_K$. 

PACS numbers: 13.10+q, 13.35.Dx, 14.40.Aq
\end{abstract}

\newpage
{\renewcommand{\thefootnote}{\fnsymbol{footnote}}


\begin{center}
D.~M.~Asner,$^{1}$ A.~Eppich,$^{1}$ J.~Gronberg,$^{1}$
T.~S.~Hill,$^{1}$ D.~J.~Lange,$^{1}$ R.~J.~Morrison,$^{1}$
R.~A.~Briere,$^{2}$
B.~H.~Behrens,$^{3}$ W.~T.~Ford,$^{3}$ A.~Gritsan,$^{3}$
% ...
\end{center}
 
\small
\begin{center}
$^{1}${University of California, Santa Barbara, California 93106}\\
$^{2}${Carnegie Mellon University, Pittsburgh, Pennsylvania 15213}\\
$^{3}${University of Colorado, Boulder, Colorado 80309-0390}\\
% ...
\end{center}

\newpage


\section{Introduction}
\label{sec:introduction}

    Decays of the $\tau$ lepton into three pseudoscalars have been
actively studied over the last several years. Lately, a number
of relatively precise measurements of the branching fractions for 
$\tau^-\to K^-\pi^+\pi^-\nu_\tau$\footnote{Charge conjugate states
are implied throughout the paper.}
 have become available
from the ALEPH, CLEO and OPAL 
collaborations\cite{aleph-97167,aleph_pre,our-paper,opal_tau96}. 
However, the resonance
substructure of these decays has not yet been measured with 
high precision. 

   The decay $\tau^-\to K^-\pi^+\pi^-\nu_\tau$,
with its simple and well-understood initial state 
provides information on 
low-$Q^2$ QCD. 
The effects of $SU(3)_f$ symmetry breaking can be observed, 
the decay constants of the $K_1$ resonances can be 
measured, and the hadronic resonance substructure can be 
studied from an analysis of the final state invariant mass 
spectra \cite{suzuki,ajw-survey}.  
Other interesting topics
include resonance parameters (such as the widths of the $K_1$ 
states), tests of isospin relations, 
and measurements of the Wess-Zumino
anomaly.
Current models of this decay 
\cite{decker,mirkes} are based on 
Chiral Perturbation Theory (ChPT) calculations.
The question of the  $K_1$ widths is of special interest because the 
theoretical  models based on ChPT \cite{mirkes} provide
 predictions for the $\tau^-\to K^-h^+h^-\nu_\tau$ 
branching fractions that are significantly
larger than current experimental values\cite{aleph_pre,our-paper}.
This discrepancy can be resolved if 
the $K_1$ resonances in $\tau$ decays are much wider than
presently measured values. 
In the non-strange sector, it has long
been realized that the $a_1$ width is considerably larger as
measured in $\tau^-\to a_1^-\nu_\tau$ compared to hadronic 
production of the $a_1$.
   The primary goal of this analysis is to measure
the relative amplitudes of the $K_1(1270)$ and
$K_1(1400)$ resonances that are believed to dominate
  $\tau^-\to K^-\pi^+\pi^-\nu_\tau$ decays\cite{decker,mirkes}
and to determine the parameters of the $K_1$ resonances.

\section{Theoretical aspects of $\tau^-\to K^-\pi^+\pi^-\nu_\tau$
decays}
\label{sec:theory}

 In the Standard Model, the general form for the 
 semileptonic $\tau$-decay matrix element can be written  
\cite{tauola} as
\begin{equation}
    {\cal M} = \frac{G}{\sqrt{2}} \bar{u}(p_\nu)\gamma^\mu(1-\gamma_5)
                        u(p_\tau)J_\mu,
\end{equation}
where $J_\mu\equiv\langle K\pi\pi|V_\mu-A_\mu|0\rangle$ is 
the hadronic current and $p_\nu$ and $p_\tau$ are the four-momenta of
the $\tau$ neutrino and the $\tau$ lepton, respectively.

General considerations based on Lorentz invariance and 
conservation of energy and momentum lead to the conclusion
that only four independent form-factors are needed to
describe the hadronic current in 
$\tau^-\to K^-\pi^+\pi^-\nu_\tau$. One parameterization\cite{decker,mirkes}
describes this process as
 \begin{eqnarray}
   J^\mu = &[F_1(s_1, s_2, Q^2)(p_2-p_3)_\nu 
          +    F_2(s_1, s_2, Q^2)(p_1-p_3)_\nu]T^{\mu\nu} \\ \nonumber
           &+     F_a(s_1, s_2, Q^2)\epsilon^{\mu\nu\rho\sigma}
                    p_{1\nu}p_{2\rho}p_{3\sigma} 
          +      F_s(s_1, s_2, Q^2)Q^\mu,        
\end{eqnarray}
where $F_i$ are form-factors, $Q^\mu$ is the 
$K\pi\pi$ 4-vector,
$s_1$ 
is expressed in terms of
the final state hadrons' momenta $p_i$  
($i$=1 for the $K^-$, $i$=2 for $\pi^+$ and
$i$=3 for $\pi^-$) as
$s_1 = (p_2+p_3)^2$, 
$s_2 = (p_1+p_3)^2$
and $T^{\mu\nu} = (g^{\mu\nu}-Q^\mu Q^\nu/Q^2)$. Here,
there are  two axial vector form-factors
$F_1$ and $F_2$, an anomalous vector form-factor $F_a$, and
a scalar form-factor $F_s$.

    To derive specific expressions for the form-factors,
some assumptions have to be made. It is believed \cite{decker,mirkes}
 that  this decay is dominated  by the lowest-mass resonances. There
are two axial vector resonances  which can produce the
$K^-\pi^+\pi^-$ final state. These are the weak eigenstates 
$^3P_1$ and $^1P_1$  $(u\bar{s})$, called $K_a$ and $K_b$. 
The $K_b$ couples
to the $W$ analagous to
a ``second class'' current, violating $SU(3)_f$ symmetry.
These two weak eigenstates mix with mixing
angle $\theta_K$ to form the observable mass eigenstates, 
$K_1(1270)$ and $K_1(1400)$ \cite{suzuki}. 
The $K_1(1270)$ subsequently
decays into $K^*\pi$, $K\rho$ or $K\rho'$,
while the $K_1(1400)$ decays almost
entirely to $K^*\pi$. 

Within the context of ChPT,
the form-factors
can be written\cite{mirkes} as
\begin{equation}
  F_1 =\frac{\sqrt{2}}{3}\cdot 
BW_{K_1(1270)}
\frac{ BW_{\rho} + 
                   \xi BW_{\rho '}}{1+\xi}
  \label{eq:f1simple}
\end{equation}
\begin{equation}
  F_2 = -\frac{\sqrt{2}}{3}\cdot 
\frac{\eta BW_{K_1(1270)}+BW_{K_1(1400)}}{(1+\eta)}
\cdot BW_{K^*} ,
  \label{eq:f2simple}
\end{equation}
where $BW$ denotes a Breit-Wigner mass distribution.
The parameter $\eta$ is estimated to be 0.33 in
 \cite{mirkes}.  The coefficients preceding the Breit-Wigner
expressions are fixed by ChPT.
In the first form-factor, the 
coefficient $\xi$ is taken to be -0.145 based on application of the
Conserved Vector Current (CVC) to $e^+e^-\to\pi^+\pi^-$ 
data \cite{mirkes,kuhn-santamaria}.

   In the chiral limit, the scalar form-factor $F_s$ is zero.
The vector form-factor $F_a$ is expected to
be numerically small compared 
to $F_1$ and $F_2$; it is only non-zero  due to 
the Wess-Zumino anomaly. 
The vector contribution is approximately $5.5\%$ as
calculated using the decay amplitudes found in \cite{tauola}.
For this analysis,
we will assume that the vector 
contribution is zero, and include our uncertainty in this term
as a systematic error.
%

ChPT has found widespread application in tau decays.
  A model similar to this is used for our analysis of the invariant
mass distributions in $\tau^-\to K^-\pi^+\pi^-\nu_\tau$,
as will be described in the following sections.

\section{Data sample and event selection}

   Our data sample contains approximately 4.3 million $\tau$-pairs
produced in $e^+e^-$ collisions, corresponding to an integrated
luminosity of 4.7 fb$^{-1}$. The data were collected with the CLEO~II
detector\cite{detector} 
at the Cornell Electron Storage Ring, operating at a center-of-mass 
energy approximately 10.58 GeV. 

The CLEO~II detector
is a general-purpose solenoidal magnet spectrometer and
calorimeter. 
The detector was
designed for efficient triggering and reconstruction of
two-photon, tau-pair, and hadronic events.
Measurements of charged particle momenta are made with
three nested coaxial drift chambers consisting of 6, 10, and 51 layers,
respectively.  These chambers fill the volume from $r$=3 cm to $r$=1 m, with
$r$ the radial coordinate relative to the beam (${\hat z}$) axis. 
This system is very efficient ($\epsilon\ge$98\%) 
for detecting tracks that have transverse momenta ($p_T$)
relative to the
beam axis greater than 200 MeV/$c$, and that are contained within the good
fiducial volume of the drift chamber ($|\cos\theta|<$0.94, with $\theta$
defined as the polar angle relative to the beam 
axis).\footnote{In this analysis we use charged tracks with 
momentum above 300 MeV/$c$.}
This system achieves a momentum resolution of $(\delta p/p)^2 =
(0.0015p)^2 + (0.005)^2$ ($p$ is the momentum, measured in GeV/$c$). 
Pulse-height measurements in the main drift chamber provide specific
ionization ($dE/dx$) resolution
of 5.5\% for Bhabha events, giving good $K/\pi$ separation for tracks with
momenta up to 700 MeV/$c$ and nearly 2$\sigma$ separation in the relativistic
rise region above 2 GeV/$c$. 
Outside the central tracking chambers are plastic
scintillation counters, which are used as a fast element in the trigger system
and also provide particle identification information from time-of-flight
measurements.  

Beyond the time-of-flight system is the electromagnetic calorimeter,
consisting of 7800 thallium-doped CsI crystals.  The central ``barrel'' region
of the calorimeter covers about 75\% of the solid angle and has an energy
resolution which is empirically found to follow:
\begin{equation}
\frac{ \sigma_{\rm E}}{E}(\%) = \frac{0.35}{E^{0.75}} + 1.9 - 0.1E;
                                \label{eq:resolution1}
\end{equation}
$E$ is the shower energy in GeV. This parameterization includes
effects such as noise, and translates to an
energy resolution of about 4\% at 100 MeV and 1.2\% at 5 GeV. Two end-cap
regions of the crystal calorimeter extend solid angle coverage to about 95\%
of $4\pi$, although energy resolution is not as good as that of the
barrel region. 
The tracking system, time of flight counters, and calorimeter
are all contained 
within a superconducting coil operated at 1.5 Tesla. 
Flux return and tracking
chambers used for muon detection are located immediately outside the coil and 
in the two end-cap regions.

    We select $e^+e^-\to\tau^+\tau^-$ 
events having a 1-prong vs. 3-prong  
topology
in which one $\tau$ lepton decays into one charged particle (plus 
possible
neutrals),
and the other
$\tau$ lepton decays into 3 charged hadrons (plus possible neutrals).
An event is separated
into two hemispheres based on the measured
event thrust axis.\footnote{The thrust axis 
of an event is chosen so that the sum of longitudinal (relative to this
axis) momenta of all charged tracks has a maximum value.} Loose
cuts on ionization measured in the drift chamber,  
energy deposited in the calorimeter and the maximum penetration
depth into the muon detector system  are applied
to charged tracks in the signal (3-prong) hemisphere to reject leptons.
Backgrounds from non-signal $\tau$ decays 
and hadronic events with $K_S^0$ are  suppressed
by requirements on the impact parameters of charged tracks.
  To reduce the background from two-photon collisions 
($e^+e^-\to e^+e^-\gamma\gamma$ with $\gamma\gamma\to$hadrons or
$\gamma\gamma\to l^+l^-$),
cuts on visible energy ($E_{vis}$) and
total event transverse momentum ($P_t$) are applied:
$2.5$ GeV $< E_{vis}< 10$ GeV, and $P_t>$0.3 GeV/$c$. We also require
the invariant mass of the tracks and showers in the 3-prong hemisphere,
calculated under the $\pi^-\pi^+\pi^-$ hypothesis,
to be less than 1.7 GeV/$c$.
Events are accepted for
which the tag hemisphere (1-prong side) is consistent with
one of the following four decays:  
$\tau^+\rightarrow e^+ \nu_e {\overline \nu_\tau}$, 
$\tau^+\rightarrow \mu^+ \nu_\mu {\overline \nu_\tau}$,  
$\tau^+\rightarrow \pi^+ {\overline  \nu_\tau}$, or
$\tau^+\rightarrow \rho^+ {\overline \nu_\tau}$. 

Candidate events are distinguished from background $\tau$ decays with 
$\pi^0$'s  and continuum hadronic background ($e^+e^-\to q\bar{q}$)
 by the characteristics of showers in the
electromagnetic calorimeter.
A photon candidate is defined as a shower in the barrel 
region of the electromagnetic calorimeter  with energy above 
100 MeV  having an energy deposition pattern consistent 
with true photons. It must be separated from the closest charged track 
by at least 30 cm.  $\tau^-\to K^-\pi^+\pi^-\nu_\tau$  
 candidates are defined as  those events having zero photon candidates 
in the 3-prong hemisphere. 

  The event selection described above provides a sample of events
that contains $\tau^-\to\pi^-\pi^+\pi^-\nu_\tau$,
$\tau^-\to K^-\pi^+\pi^-\nu_\tau$ and $\tau^-\to K^-K^+\pi^-\nu_\tau$.
In this analysis we neglect possible contributions from the decays 
$\tau^-\rightarrow \pi^-K^+\pi^-\nu_\tau$ and  
$\tau^-\rightarrow K^-\pi^+K^-\nu_\tau$ because they
are unphysical in the 
Standard Model and have not been experimentally observed.
We also neglect the  $\tau^-\rightarrow K^-K^+K^-\nu_\tau$ 
final state. 
This rate is expected to be $\sim 1\%$ relative to that for
$\tau^-\rightarrow K^-\pi^+\pi^-\nu_\tau$ due to the limited
phase space and the low probability of 
forming an $(s\bar{s})$ pair from the vacuum.

\section{ Reconstruction of invariant mass spectra}

Due to the very small fraction of kaons in $\tau^-\to h^-h^+h^-\nu_\tau$
events\footnote{Here and later $h$ designates either a kaon or pion.}
 and the limited particle identification 
capabilities of the CLEO~II detector,
it is difficult to
identify individual $\tau^-\to K^-\pi^+\pi^-\nu_\tau$ decays.
In this analysis, a statistical approach is used in which the number
of $\tau^- \to K^-h^+\pi^-\nu_\tau$  events in any given sample is 
determined using the $dE/dx$ information of the two 
same-sign tracks in the signal hemisphere. The $dE/dx$ analysis 
 is described in detail in \cite{our-paper}.

   For each $h^-h^+h^-$ candidate the invariant mass of the three 
hadrons is calculated under two hypotheses for the first
and the third tracks, corresponding to the $K^-\pi^+\pi^-$ and
$\pi^-\pi^+K^-$ mass assignments.  
Each of these two sub-samples is divided into bins
of invariant mass. The bins are 100 MeV/$c^2$
wide, spanning  the region $0.8-1.7$ GeV/$c^2$. After binning
in mass, the sub-samples  that
correspond to the same mass bin are combined.  

   The  $dE/dx$ analysis provides the number of kaons in each mass bin,
which is equal to the number of  $\tau^- \to K^-h^+\pi^-\nu_\tau$ 
events in that mass interval. The invariant mass spectrum 
of the $K^-\pi^+\pi^-$ system is thereby reconstructed. 
This distribution contains a contribution from  
$\tau^- \to K^-K^+\pi^-\nu_\tau$ decays which must be subtracted, as will be 
discussed in Sec.\ref{sec:backgrounds}.

  In a similar way the invariant mass spectra of $K^-\pi^+$ and
$\pi^+\pi^-$ are reconstructed in ten bins
over the range $0.5$ GeV/$c^2<M_{K^-\pi^+}<1.5$ GeV/$c^2$ and 
$0.2$ GeV/$c^2<M_{\pi^+\pi^-}<1.2$ GeV/$c^2$, respectively. 
The reconstructed mass spectra are shown in Fig. \ref{mass}.

\begin{figure}
\vspace{11cm}
    \begin{minipage}[t]{7cm}
\includegraphics{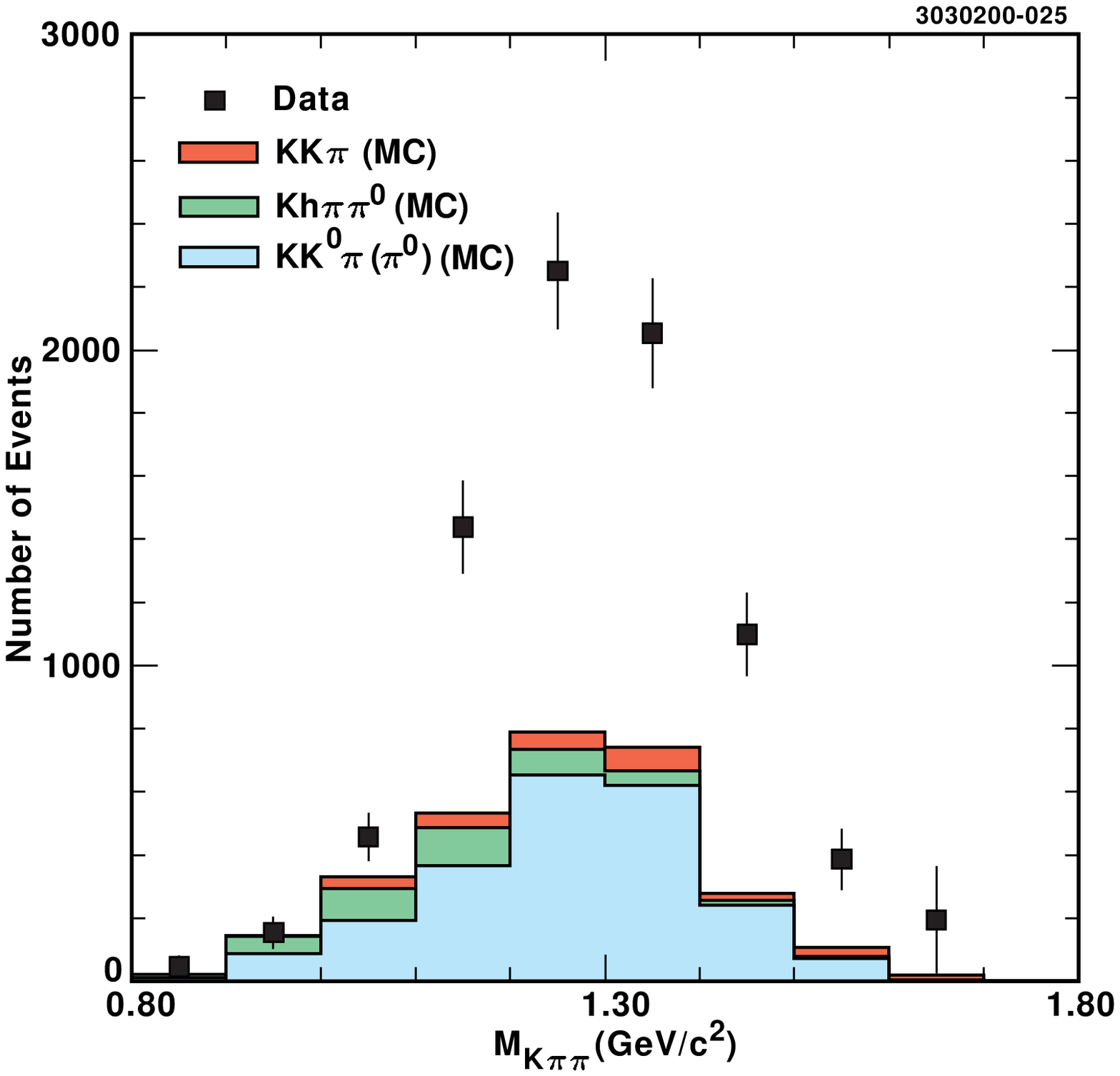}
   \end{minipage}
   \hfill
   \begin{minipage}[t]{7cm}
\includegraphics{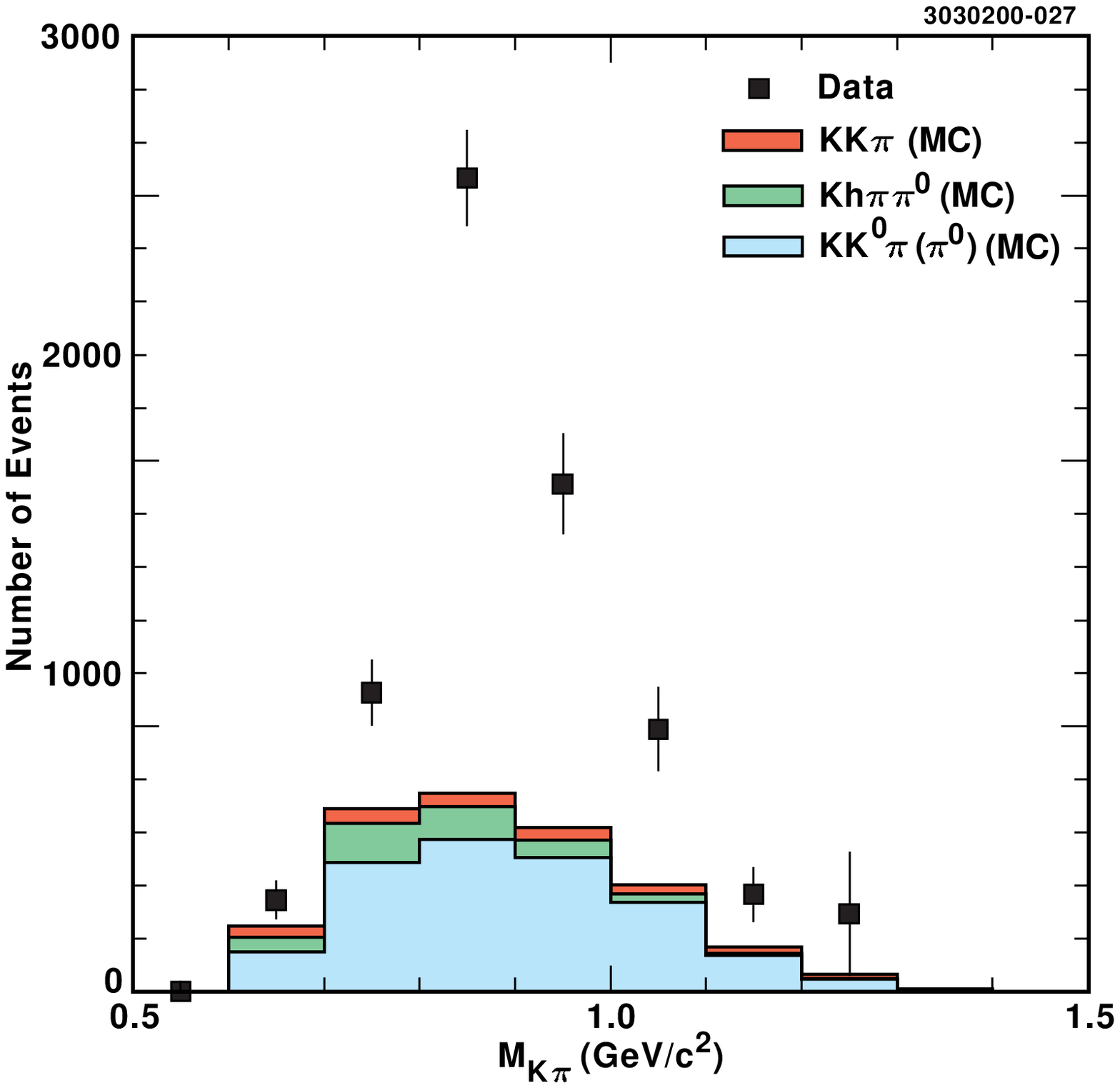}
   \end{minipage}

\vspace{11cm}
\includegraphics{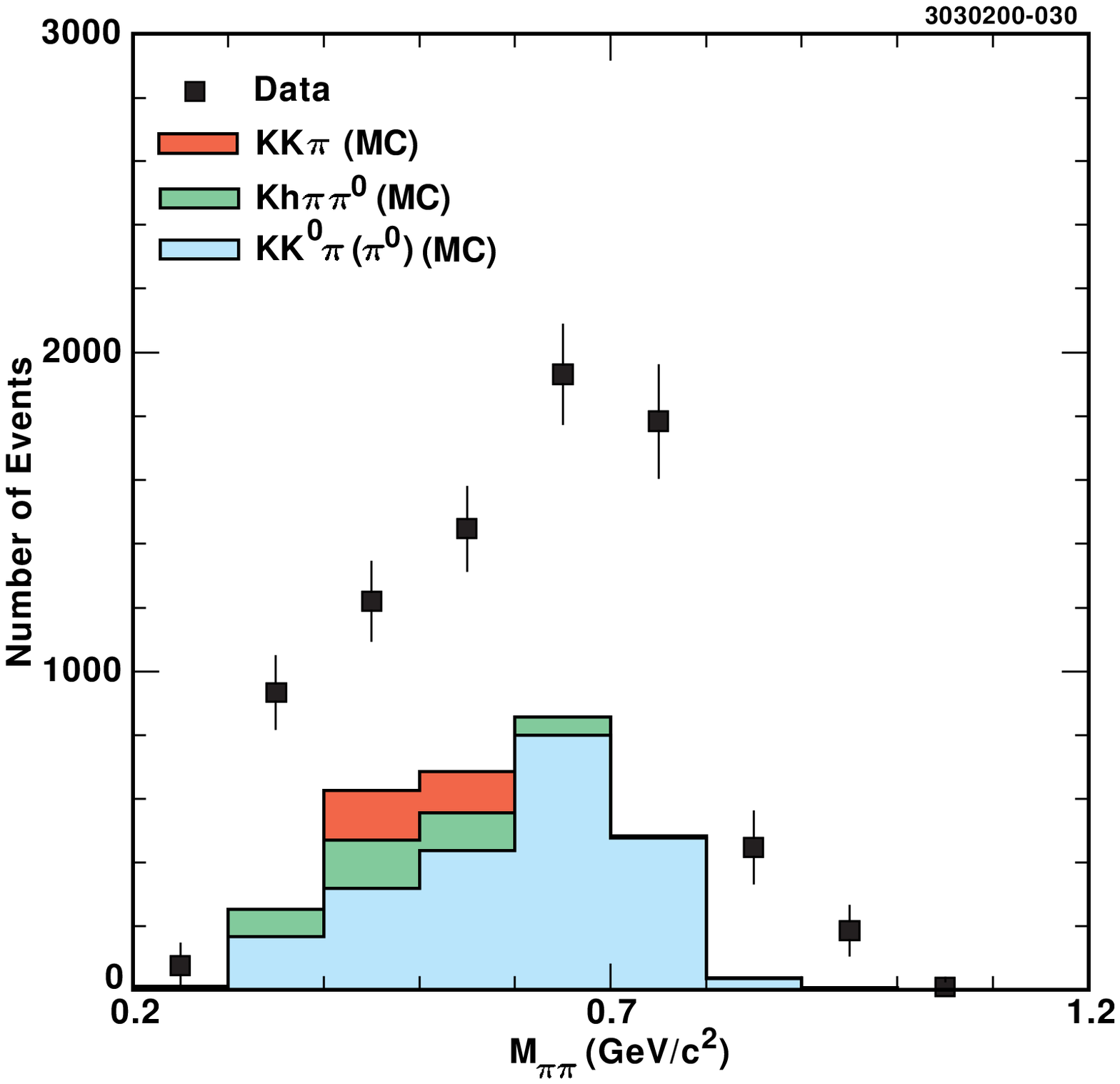}
         \caption{Reconstructed mass spectra (data) and 
backgrounds predicted from Monte Carlo simulations. 
                    \label{mass}}
\end{figure}

\section{Background and efficiency}
\label{sec:backgrounds}

   There are two main types of background: continuum hadronic
events ($e^+e^-\to q\bar{q}\to hadrons$) 
and non-signal $\tau$ decays. Hadronic background is estimated
from a continuum hadronic Monte Carlo sample (using the JETSET v7.3
\cite{JETSET} event generator and GEANT\cite{GEANT}
detector simulation code). 
This background is subtracted as described in 
\cite{our-paper}. The level of hadronic background is 
approximately 3\%.

 $\tau$-related background comes primarily from
$\tau^-\to K^-K^+\pi^-\nu_\tau$ decays. These events
comprise approximately $27\%$ of the events in our reconstructed 
invariant mass distributions. Smaller background contributions arise
from $\tau^-\to K^-h^+\pi^-\pi^0\nu_\tau$ with incomplete 
$\pi^0$ reconstruction and also tau
decays to the  $K^-\pi^+\pi^-(\pi^0)\nu_\tau$ final state through
an intermediate $K^0_S$.
These two backgrounds comprise 5\% and 3\% of the
events in the $K^-\pi^+\pi^-$  invariant mass spectra, 
respectively. The invariant mass distributions
for backgrounds are found using Monte Carlo simulations
to obtain the shape;
the normalization is set by the measured branching fractions 
\cite{our-paper,PDG98}.
Background  predictions are shown in Fig. \ref{mass}.
The invariant mass distributions for all backgrounds
are subtracted from the corresponding invariant mass spectra
reconstructed from data. 

The efficiency of event reconstruction depends slightly on
the invariant mass. Therefore, it is necessary to introduce
a mass-dependent efficiency correction. This correction is calculated
from $\tau$ Monte Carlo using the KORALB event generator \cite{tauola}.
The maximum variation in efficiency across the mass interval
of interest is of order 10\%.

\section{Fitting method}
\label{sec:our-model}   

    The hadronic structure of the $K^-\pi^+\pi^-$ system 
is investigated by simultaneously fitting three invariant mass
distributions: $M_{K^-\pi^+\pi^-}$, $M_{K^-\pi^+}$ and $M_{\pi^+\pi^-}$. 
The fitting function is based upon a model similar to the one 
described in  Sec. \ref{sec:introduction}, and now outlined
in greater detail.

\subsection{Parameterization of form-factors}



In this analysis,
we write the following expression for the axial vector form-factors
$F_1$ and $F_2$:

\begin{equation}
   F_1 = \frac{1}{\sqrt{3}}(A\cdot BW_{1270}
           + B\cdot BW_{1400})
\frac{ BW_{\rho} + 
                   \xi BW_{\rho '}}{1+\xi}
 \label{eq:f1new}
\end{equation}
\begin{equation}
   F_2 = (-\frac{2}{3})(C\cdot BW_{1270} + D\cdot BW_{1400})
              BW_{K^*}   
 \label{eq:f2new}
\end{equation}
%
%
that
contain the four real parameters $A-D$.
Of these, 3 are independent; the fourth is fixed by the 
normalization requirement that the squared
sum of the $\tau^-\to K_1^-(1270)\nu_\tau$ and
$\tau^-\to K_1^-(1400)\nu_\tau$ amplitudes must saturate the total 
$\tau^-\to K^-\pi^+\pi^-\nu_\tau$
rate. The coefficients $A$ and $B$ correspond to production of the
$K\rho$ final state through either $K_1$(1270) ($A$) or $K_1$(1400) ($B$),
modulo a factor which includes the appropriate phase space weighting
for various final states (denoted as ``$R_A$'', or ``$R_B$''). In our
analysis, we fix $B$ to be zero, consistent with current
measurements \cite{PDG98}.
Similarly, $C$ and $D$ designate production of the $K^*\pi$ final state
through the $K_1(1270)$ and $K_1(1400)$ resonances.
The decay amplitude parameters in Eqs. (\ref{eq:f1new})-(\ref{eq:f2new})
therefore correspond to the possible decay chains as
\begin{eqnarray}
  \tau\to K_1(1270)\nu\to K\rho\nu\to K^-\pi^+\pi^-\nu_\tau:&
``A'' &
      \label{eq:defA}\\
  \tau\to K_1(1270)\nu\to K^*\pi\nu\to K^-\pi^+\pi^-\nu_\tau:&
``C''=& 
       A\cdot\sqrt{\frac{16}{42}}\cdot{\sqrt{\frac{R_A}{R_C}}}
      \label{eq:defC}\\
  \tau\to K_1(1400)\nu\to K^*\pi\nu\to K^-\pi^+\pi^-\nu_\tau:& 
``D'' =&\sqrt{1-A^2-C^2}               
      \label{eq:defD}  
\end{eqnarray}
In Eqs. (\ref{eq:defA})-(\ref{eq:defC}) we have imposed constraints
that follow from  the tabulated branching fractions of the 
$K_1$ resonances \cite{PDG98}:
 ${\cal B} ({K_1(1270)\to K^*\pi}) = (16\pm 5)$\%~and 
 ${\cal B} ({K_1(1270)\to K\rho} )= (42\pm 6)$\%. 

 Thus, in our parameterization of the matrix element one
unknown parameter $A$ defines all four amplitudes. In addition, the
masses and widths of the
$K_1$ resonances $\Gamma_{K_1(1270)}$, $\Gamma_{K_1(1400)}$,
$M_{K_1(1270)}$, $M_{K_1(1400)}$ are considered unknown and left
as free parameters in the fit. The Breit-Wigner
distributions for the $K_1$ resonances are defined following the
approach of \cite{tauola} as
\begin{equation}
   BW(s,m_{K_1}, \Gamma_{K_1}) = 
      \frac{m_{K_1}^2 - im_{K_1}\Gamma_{K_1}}
         {m_{K_1}^2-s-im_{K_1}\Gamma_{K_1}}\;\;.
   \label{eq:bw1}  
\end{equation}
The Breit-Wigner distributions for the $K^*$ and $\rho$ resonances contain
 mass-dependent widths:
\begin{equation}
   BW(s,m, \Gamma) = 
      \frac{m^2}
         {m^2-s-i\sqrt{s}\Gamma(s)},
   \label{eq:bw2}  
\end{equation}
where the mass-dependence is defined by Eq. (22) in \cite{tauola}:
\begin{equation}
    \Gamma (s) = \Gamma_0 \frac{m_0^2}{s}
              \left( \frac{p(s)}{p(m_0^2)} \right)^\frac{3}{2}.
\label{eq:mass-dependent-width}
\end{equation}
Here, $m_0$ and $\Gamma_0$ are the nominal mass and width of a particle,
$p$ is the momentum of the particle, and 
the variable $s$
is the three- or two-body invariant mass, as appropriate.

  The constants $R_X$ (where $X=A,...,D$) in Eqs. 
(\ref{eq:defA}), (\ref{eq:defC}), and (\ref{eq:defD})
depend on the masses and widths of the 2- and 3-body 
resonances in this decay and are calculated  by numerical integration
of the appropriate matrix element. Since we 
are interested in the ratios of quantities (e.g.,
branching fractions), the overall normalization of the $R_X$ parameters
is arbitrary.

In this analysis, we have  
determined the numerical coefficients for the Breit-Wigner terms 
[$\sqrt{\frac{1}{3}}$ and $-\frac{2}{3}$ in Eqs. 
(\ref{eq:f1new})-(\ref{eq:f2new})] 
using isospin relations rather than taking the chiral limit as in 
\cite{mirkes}. 
  The Wess-Zumino anomaly term is set to zero in our model;
this term is numerically small enough that it can be neglected at 
our level of  accuracy. Appropriate systematic errors are assigned to 
reflect the possible magnitude of this contribution.
Note,
however, that if the Wess-Zumino anomaly is much larger than expected, 
the $K^{*'}$ may contribute events to the region of $M_{K^-\pi^+\pi^-}$
invariant mass close to 1.4 GeV/$c^2$, affecting our measurement 
of $A-D$.  Note also that we  explicitly
assume  all $\tau^-\to K^-\pi^+\pi^-\nu_\tau$ decays proceed
through either $K_1(1270)$ or $K_1(1400)$.

In principle,
there may be  a phase shift between the terms in 
the form-factors $F_1$ and $F_2$.
Such phase differences may appear between various decay
chains producing the final state $K^-\pi^+\pi^-$. This
may cause additional constructive or destructive interference
and, for example, enhance or suppress the $\rho$ peak in 
the distribution of $\pi\pi$
invariant mass. 
In the most general approach,
one would introduce three independent phase angles $\theta_1$,   
$\theta_2$ and $\theta_3$, corresponding
to the possible interfering decay chains.
However, due to limited statistics, we have neglected such possible 
interference effects, and take into account only the inherent phase of the
Breit-Wigner distributions (as described above).

\subsection{Calculation of the observables}

  The interesting observables that we would like to measure are
the relative branching fractions to the different $K_1$ resonances and 
the amounts
of $K^*$ and $\rho$ in this decay. The decay rate for any individual
decay chain is proportional to $X^2R_X$. For example, the decay rate
for the chain in Eq. \ref{eq:defA} is proportional to $A^2R_A$. In calculating
the ratios we choose to normalize to the sum of the separate contributions
(not including interference effects). With this convention, 
the fractions of different contributions 
add to 100\%.

   With the above definitions we write
\begin{equation}
   f_{1270}\equiv \frac{{\cal B}(\tau\to K_1(1270)\nu)}
     {{\cal B}(\tau\to K_1(1270)\nu) + {\cal B}(\tau\to K_1(1400)\nu)} 
                  = \frac{A^2R_A + C^2R_C}
                                       {A^2R_A + C^2R_C + D^2R_D}
\label{eq:rate1270}
\end{equation}
and 
\begin{equation}
  f_{\rho} \equiv
    \frac{{\cal B}(\tau\to K\rho\nu)}
         {{\cal B}(\tau\to K\pi\pi\nu)} = \frac{A^2R_A}
                                {A^2R_A + C^2R_C + D^2R_D}.
\label{eq:rateRho}
\end{equation}

\subsection{Fitting function}

  We use a Monte Carlo based fitting procedure, in which 
a large number (200,000) 
of simulated events are used to simultaneously fit 
the data distributions for $M_{K\pi\pi}$, $M_{K\pi}$ and  
$M_{\pi\pi}$. 
From a binned $\chi^2$ fit,
we determine the input values of $A$, 
$\Gamma_{K_1(1270)}$, $\Gamma_{K_1(1400)}$,
$M_{K_1(1270)}$ and $M_{K_1(1400)}$ which give the best 
simultaneous match to these 
mass spectra.
As outlined above,
our event generator is identical 
to KORALB \cite{tauola} except
that our form-factors [Eqs.(\ref{eq:f1new})-(\ref{eq:f2new})]
are used and the Wess-Zumino form-factor is set to zero. 
To take into
account finite resolution effects we introduce Gaussian
smearing of the calculated mass equal to the smearing found 
from the full GEANT-based \cite{GEANT} simulation of the detector. 
This smearing is typically 5-10 MeV/$c^2$.

\section{Fit results}

\label{sec:analysisa}


The result of our fit is shown in 
Figure \ref{fit1}; the best values for the fit parameters
 are tabulated in Table \ref{tab:full-fit}. In the same
table the values of $R_X$ obtained from numerical integration
and the derived values for $f_{1270}$ and $f_\rho$ are also given. 
The first error in the Table is statistical and the second
is systematic (discussed in 
Sec. \ref{sec:syst}).
The statistical errors on $A$, the masses and widths are
calculated using HESSE in MINUIT\cite{MINUIT} and take into account
correlations between the fit parameters.
The asymmetric statistical errors, where
appropriate, are also evaluated from the fit.

\begin{figure}
\vspace{10cm}
   \begin{minipage}[t]{4cm}
 \includegraphics{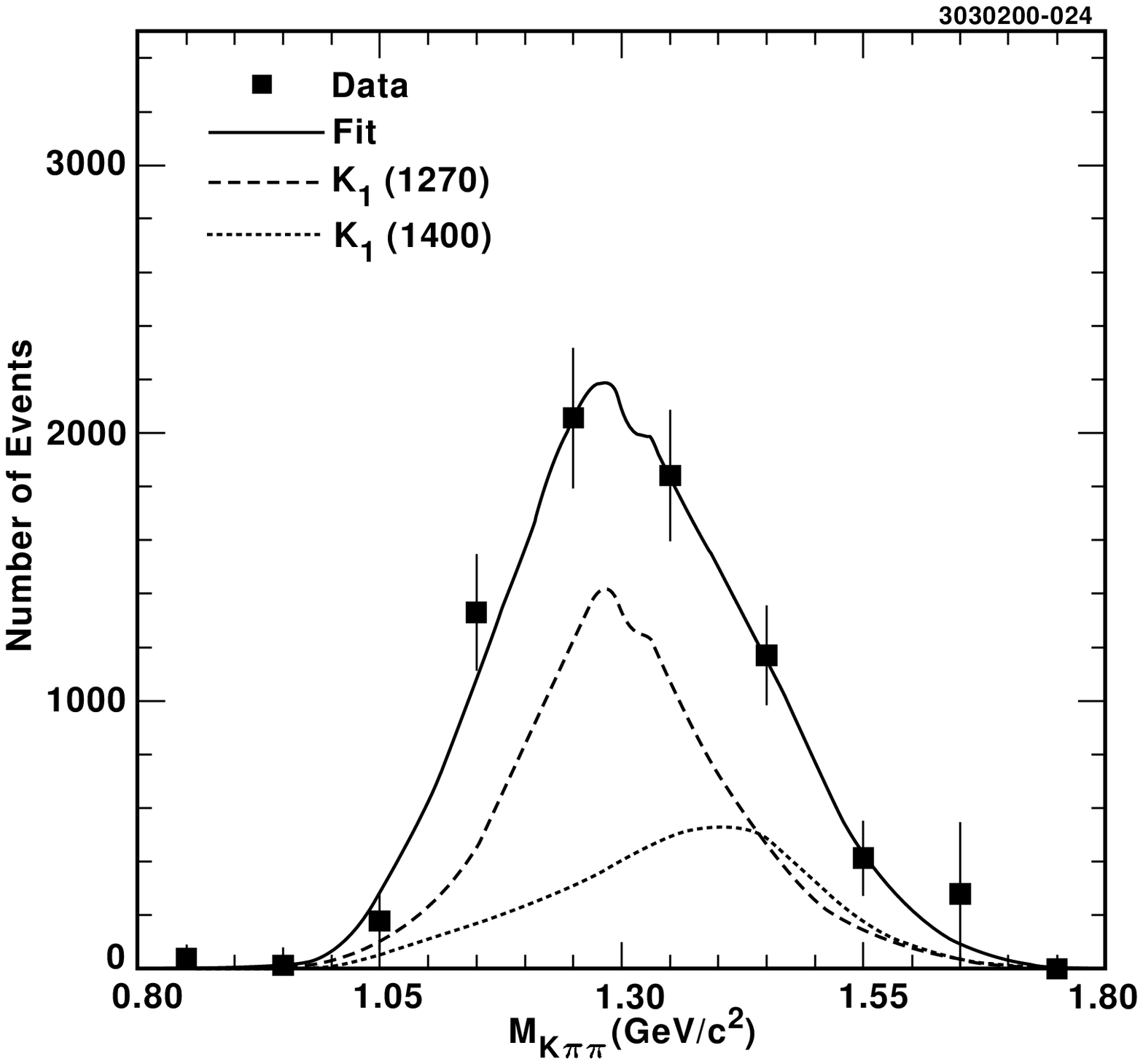}
   \end{minipage}
   \hfill
   \begin{minipage}[t]{4cm}
 \includegraphics{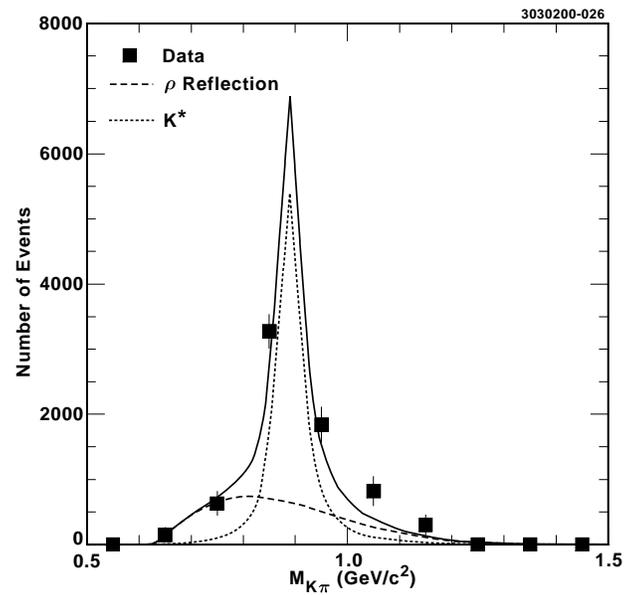}
   \end{minipage}

\vspace{10cm}
 \includegraphics{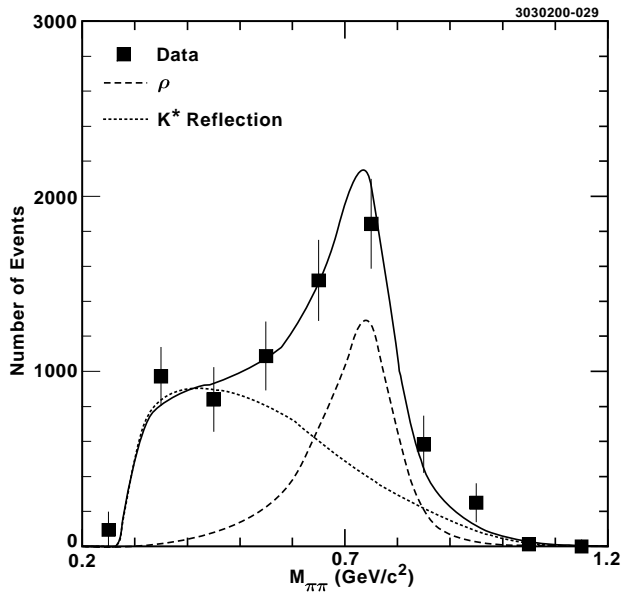}
\caption[]{Fit to the distributions of $M_{K\pi\pi}$, 
$M_{K\pi}$ and $M_{\pi\pi}$ with individual components overlaid,
after subtracting the
non-signal backgrounds in Fig. 1, and after correcting for the
relative efficiency as a function of mass.
In the upper left plot, the
solid line represents the sum of the $\tau\to K_1(1270)\nu_\tau$ + 
$\tau\to K_1(1400)\nu_\tau$ mass spectra
and the dashed and dotted lines represent the individual
$K_1$(1270) and $K_1$(1400) contributions. 
Similarly, the upper right
and bottom plots show the $K^*\pi$ contributions (dotted) and the 
$K\rho$ contributions (dashed) to the $K\pi\pi$ final state.
The limited statistics of the fitting function is evident in the
non-smoothness of the solid curve in these three figures.
\label{fit1}}
\end{figure}

\begin{table}
\caption[]{Results for the full fit. 
In the fit, the parameter $B$ is set to zero and the parameters
$C$ and $D$ are constrained by the branching fractions\cite{PDG98} into
the $K^*\pi$ final state from the $K_1(1270)$ and $K_1(1400)$, 
respectively. The parameters $R_A$, $R_B$, $R_C$ and $R_D$ are 
normalizations returned from the fit, as described in the text. Errors
for the fit parameters $M$, $\Gamma$ and $f$ are statistical and
systematic (respectively), as described in the text.
\label{tab:full-fit}}
\center
\begin{tabular}{ll}
$A=0.94\pm 0.03\pm 0.04$ & $R_A=6917$\\
$B=0.00$ (fixed) & $R_B=12403$\\
$C=0.20$ (constrained)& $R_C=61636$  \\
$D=0.27$ (constrained) & $R_D=58027$\\
$\Gamma_{1270}=0.26^{+0.09}_{-0.07}\pm 0.08$ GeV & \\
$\Gamma_{1400}=0.30^{+0.37}_{-0.11}\pm 0.14$ GeV 
                            &$f_{1270} = 0.66\pm 0.19\pm 0.13$ \\  
$M_{1270}=1.254\pm 0.033 \pm 0.034$ GeV/$c^2$ 
                            & $f_\rho = 0.48\pm 0.14\pm 0.10$ \\
$M_{1400}=1.463\pm 0.064 \pm 0.068$ GeV/$c^2$ & \\
\end{tabular}
\end{table}


The fit results showing
contours of constant $\chi^2$
in the $M_{1270}$ vs.
$M_{1400}$ and $\Gamma_{1270}$ vs. $\Gamma_{1400}$ planes
are shown in Figures
\ref{kab_wmass} and \ref{kab_width}.
(Note that the constraint $M_{1400}>M_{1270}$ is introduced in obtaining
Figures \ref{kab_wmass}-\ref{kab_width}; we therefore fit only over
the region above the dashed line.) In these plots, the curves
represent 1$\sigma$, 2$\sigma$, 3$\sigma$ and 4$\sigma$ 
standard deviation error
contours around the best fit point (indicated by a cross).

\begin{figure}
\vspace{10cm}
\includegraphics{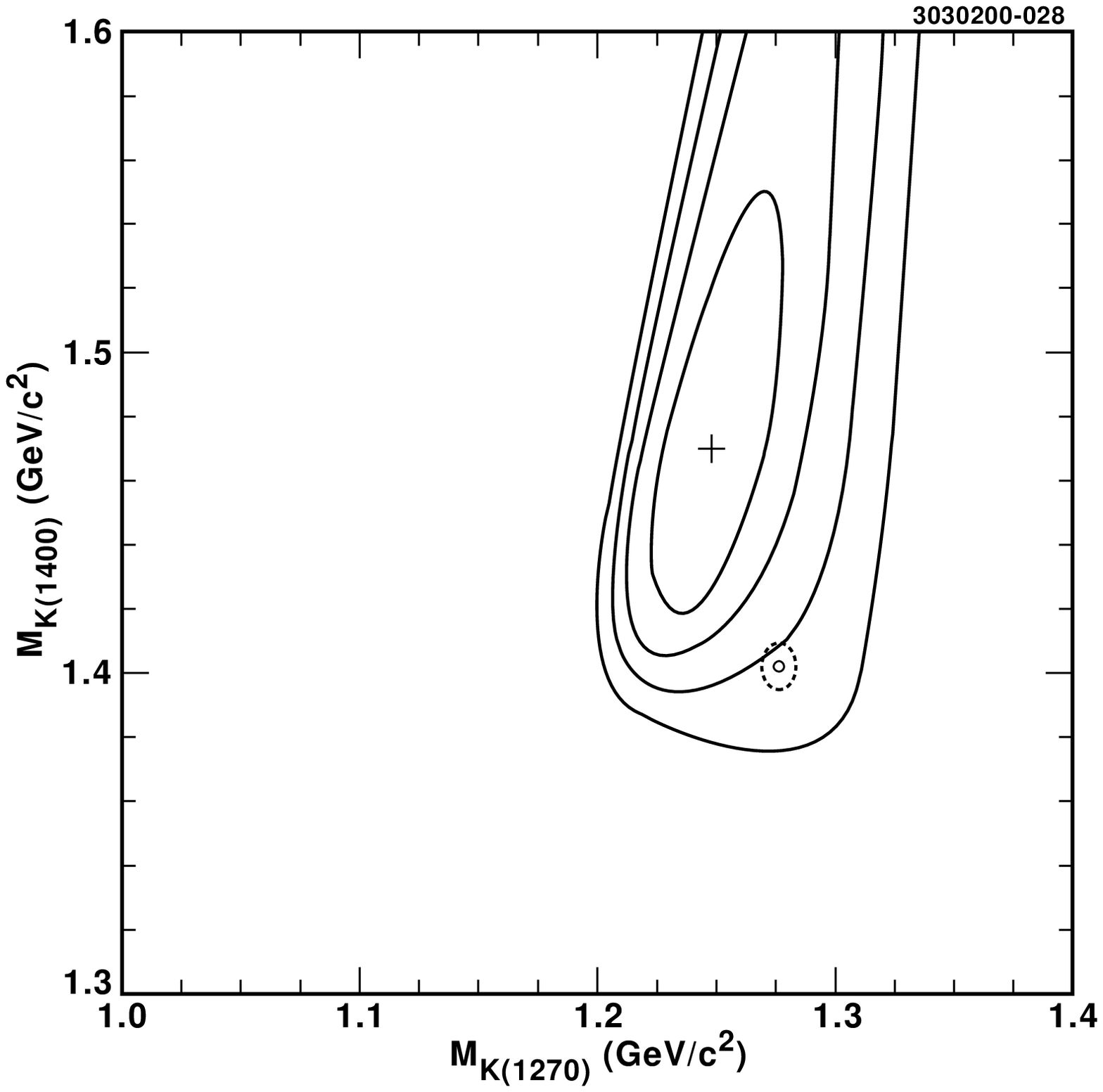}
\caption[]{Scan of the mass parameter plane  
showing fit results for the $K_1$(1270) mass
(horizontal) vs. the $K_1$(1400) mass (vertical).
Also shown are 1-4 $\sigma$ standard deviation error
contours (statistical errors only) around our best fit value (cross)
The small open circle shows the present PDG values for the $K_1$(1270) and 
$K_1$(1400) masses,
with the associated errors (dotted ellipse). The systematic errors
for our measurement,
although not shown, are nevertheless significant (see text). 
\label{kab_wmass}}
\end{figure}

\begin{figure}
\vspace{10cm}
   \includegraphics{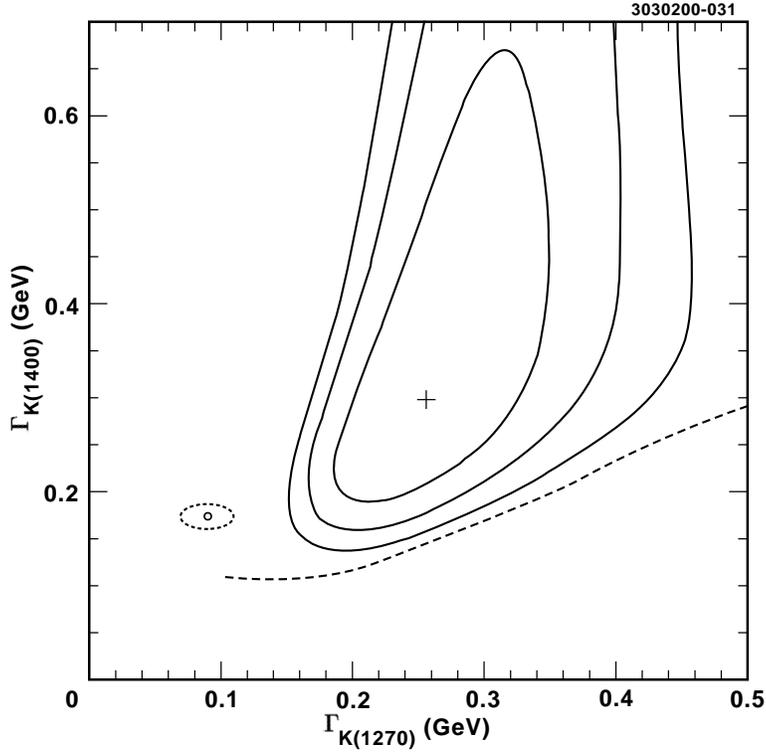}
\caption[]{Scan of the width parameter plane 
showing fit results for the $K_1$(1270) width
(horizontal) vs. the $K_1$(1400) width (vertical), as well
as contours of constant
$\chi^2$
(1-3 $\sigma$, statistical errors only).
The small open
circle shows the present PDG values for the $K_1$(1270) and 
$K_1$(1400) widths,
with the associated errors (dotted ellipse).
The dashed line results from the fitting condition
that $M_{K_1(1270)}<M_{K_1(1400)}$.
\label{kab_width}}
\end{figure}

We perform a 
secondary fit, in which the
masses and widths of the $K_1$ resonances are fixed
to world
average values \cite{PDG98}.
We obtain $f_{1270}=0.40\pm 0.07$, and $f_{\rho}=0.29\pm 0.05$ from
this second fit (statistical errors only), to
be contrasted with the significantly larger values extracted from our
full fit, in which the $K_1$ masses and widths are allowed to float as free
 parameters. It is not surprising that the $f$ values are different in this
second fit compared to the original fit, 
given the high degree of
correlation between the $K_1$ widths and the relative branching fractions.
The $\chi^2$ for this second fit is considerably poorer 
(30.5/22 degrees of freedom) than the primary fit
(12.6/18 degrees of freedom).

\section{Systematic uncertainties}
\label{sec:syst}

  Systematic errors are summarized in Table \ref{tab:sys-full}.
The dominant errors are due to the uncertainty in the $K_1$
branching fractions and the uncertainty in 
the background.
The errors associated with the uncertainty in the 
branching fractions of the $K_1$ resonances 
to $K^*\pi$ and $K\rho$ are estimated
by changing the values for these branching fractions in the
form-factors $F_1$ and $F_2$ by one standard deviation of the 
PDG values\cite{PDG98}. The resulting spread in the fit values 
 is taken as the corresponding systematic error. 

\begin{table}[hbtp]
\caption[]{\label{tab:sys-full} Summary of systematic 
error calculations for the full fit. ``$K_1$ parameters'' refers to the 
uncertainty in the branching fractions, ``Model dependence'' includes the
uncertainty in the 
kaon momentum spectrum
in $\tau^-\to K^-h^+h^-\nu_\tau$ ($P_K$ spectrum).
Due to the correlation between momentum and three-prong mass,
coupled with the fact
that we have good particle identification only over a limited momentum
range, uncertainty in $P_K$ results in a corresponding uncertainty in the
mass spectrum. Other errors are as indicated.}
\center
\begin{tabular}{lllllllll}
\multicolumn{2}{l}{ Source}              
& $\Gamma_{1270}$ & $\Gamma_{1400}$ 
                    & $M_{1270}$ & $M_{1400}$ 
& $A$ & $f_{1270}$ & $f_\rho$ \\  
\multicolumn{2}{l}{ }              
& GeV & GeV & GeV/$c^2$ & GeV/$c^2$ & & & \\
\hline
\multicolumn{2}{l}{$K_1$ parameters} 
  & 0.005 & 0.049 & 0.005 & 0.004    & 0.011 & 0.016 & 0.012 \\
\hline
Model 
dependence    &$P_K$ spectrum          
  & 0.057 & 0.038 & 0.004 & 0.020 & 0.012 & 0.075 & 0.055 \\
    &$M_{KK\pi}$ shape                 
  & 0.039 & 0.107 & 0.024 & 0.055 & 0.010 & 0.083 & 0.060 \\
    &$\rho^\prime$ contribution                    
  & 0.006 & 0.001 & 0.005 & 0.003 & 0.012 & 0.005 & 0.004 \\
    &Vector current                         
  & 0.015 & 0.026 & 0.009 & 0.005 & 0.005 & 0.025 & 0.025 \\
\hline
\multicolumn{2}{l}{Background level} 
  & 0.017 & 0.054 & 0.020 & 0.033 & 0.028 & 0.054 & 0.039 \\ 
\hline
\multicolumn{2}{l}{Function and MC statistics} 
  & 0.019 & 0.026 & 0.005 & 0.006 & 0.006 & 0.026 & 0.019 \\
\hline
\hline
\multicolumn{2}{l}{Total systematic error} 
  & 0.076 & 0.140 & 0.034 & 0.068 & 0.037 & 0.130 & 0.097 \\ 
\end{tabular}
\end{table}

   The uncertainty in the background is large because of the uncertainty
in the branching fraction of its largest component, 
$\tau^-\to K^-K^+\pi^-\nu_\tau$. During the background subtraction
the level of all $\tau$-related backgrounds are
 varied by amounts corresponding to
the errors on the branching fractions of these decays 
\cite{our-paper,PDG98}.
The hadronic background is similarly varied by 100\% to determine
the systematic error due to our uncertainty in the $q\bar{q}$ 
background contribution.

   Another large error comes from the choice of models in our Monte Carlo
simulation. This includes the uncertainty in the shape of the kaon 
momentum spectrum ($P_K$) used to obtain the total number of events with kaons
\cite{our-paper}, and the uncertainty in the shape of the invariant mass 
distribution for the $\tau^-\to K^-K^+\pi^-\nu_\tau$ background.
These errors are estimated by using several different models
to extract the invariant mass spectra.
For $\tau^-\to K^-\pi^+\pi^-\nu_\tau$,
we consider
$\tau^-\to K_1(1270)^-\nu_\tau$ and $\tau^-\to K_1(1400)^-\nu_\tau$
and the model described in \cite{tauola};  
for $\tau^-\to K^-K^+\pi^-\nu_\tau$, we consider
$\tau^-\to a^-_1\nu_\tau \to K^{*0}K^-\nu_\tau$,
$\tau^-\to\rho(1690)^-\nu_\tau \to K^{*0} K^-\nu_\tau$ 
and the model described in \cite{tauola}.


There are several fitting function uncertainties. The first
is the contribution from
the $\rho'$ which may be different in this decay from that
observed in $e^+e^-\to \pi^+\pi^-$ data \cite{kuhn-santamaria}
due to the phase space suppression of 
$\rho'$ in our case.  Second,  the model implemented in our fitting 
function contains no contribution from the vector current.
The corresponding fitting function errors from these two sources are 
estimated by varying the level
of the Wess-Zumino term and the $\rho'$ amplitudes from zero
to the predictions of \cite{mirkes} and \cite{li}. 
Another possible source of systematic errors is a phase
shift among the interfering decay chains. In this
analysis, the parameters $A-D$
in Eqns. (\ref{eq:f1new})-(\ref{eq:f2new}) are real.
We have done a study of interference effects with additional phase
shifts and found that the possible imaginary part of $A-D$
is consistent with zero at our level of sensitivity.
 Because the fitting function is based upon Monte Carlo, the Monte Carlo 
statistical error is also included here.

   The bias associated with the procedure of fitting 
the invariant  mass distribution 
is studied using 60 samples of signal Monte Carlo
with a full detector simulation. The results of this study 
show no systematic shift of the fitted parameter values relative
to the input values.
Additionally, the errors we obtain from analyzing this Monte Carlo
sample are fully consistent with statistical expectations.

In this analysis  only the shape of the background-subtracted
invariant mass
distribution is of interest; possible systematic effects
that affect the overall normalization  of the reconstructed spectra
are ignored. Among such effects are trigger and tracking efficiencies,
and the photon veto.
Non-$\tau$ 
backgrounds (2-photon events, beam-gas interactions, QED background, e.g.)
have been determined to be negligible for the mass spectrum analysis.

\section{Discussion of Resonance Structure}
\subsection{Masses and Widths}

As mentioned previously, theoretical predictions for
${\cal B}(\tau\to Kh^+h^-\nu_\tau)$ 
based on ChPT\cite{mirkes} are substantially
larger than data. However, if the $K_1$ resonances are substantially
broader than the PDG values, this discrepancy is resolved. In fact,
our data suggest larger $K_1$ widths
than previous world averages \cite{PDG98} (this is evident from 
Fig. \ref{fit1}). As indicated in Table \ref{tab:full-fit},
we extract the masses and widths of the $K_1$ resonances from this
fit: $\Gamma_{1270} = 0.26^{+0.09}_{-0.07}$ GeV,
$\Gamma_{1400} = 0.30^{+0.37}_{-0.11}$ GeV,
$M_{1270} = 1.254\pm 0.033$ GeV/$c^2$, and
$M_{1400} = 1.463\pm 0.064$ GeV/$c^2$. 

  In Table \ref{tab:widths},
our result for the widths is compared to the 
data from ALEPH and DELPHI \cite{mirkwidth}
in their analyses of $\tau$ decays. One observes that all experimental
data from $\tau\to K\pi\pi\nu_\tau$ for the
$K_1$ widths are above the current world
averages although the errors remain large.
  The masses of $K_1(1270)$ and $K_1(1400)$ measured in our
analysis are in acceptable agreement with the  current world averages 
\cite{PDG98}.

\begin{table}
\caption[]{\label{tab:widths} Current world average and measurements
for $\Gamma_{K_1}$ from $\tau\to K\pi\pi\nu_\tau$.}
\center
\begin{tabular}{lll}
          &  $\Gamma_{K_1(1270)}$, GeV 
                           & $\Gamma_{K_1(1400)}$, GeV \\
\hline 
 PDG \cite{PDG98}
    &      $0.09\pm 0.02$         &     $0.174\pm 0.013$  \\
\hline
 ALEPH \cite{mirkwidth}
    &     $0.37\pm 0.10$        &    $0.63\pm 0.12$ \\
 DELPHI \cite{mirkwidth}
    &     $0.19\pm 0.07 $        &    $0.31\pm 0.08 $ \\
 this analysis & $0.26^{+0.09}_{-0.07}\pm 0.08$        
                       &     $0.30^{+0.37}_{-0.11}\pm 0.14$ \\
\end{tabular}
\end{table}

\subsection{Values of $f_{1270}$ and $f_\rho$}

As calculated in section VII,
 our data indicate that there is slightly more $K_1(1270)$
than $K_1(1400)$ in the axial vector current of the
$\tau^-\to K^-\pi^+\pi^-\nu_\tau$ decay 
($f_{1270} = 0.66\pm 0.19\pm0.13$). 
Other experiments have also investigated the 
relative contributions of the two $K_1$ resonances to 
$\tau\to s\overline{u}\nu_{\tau}$.
One of the first measurements of the $\tau\to K_1\nu_\tau$ branching
fractions was performed by the $TPC/2\gamma$ collaboration in 1994 \cite{tpc}. 
Their
results are ${\cal B}(\tau\to K_1(1270)\nu)=0.41^{+0.41}_{-0.35}$\%
and ${\cal B}(\tau\to K_1(1400)\nu)=0.76^{+0.40}_{-0.33}$\%, giving
the fraction $f_{1270} = 0.35^{+0.73}_{-0.35}$. The results
of the $TPC/2\gamma$ experiment suggest that the decay proceeds mostly 
through $K_1(1400)$ although their errors are too large to draw 
firm conclusions.
 The latest branching fraction
measurements by CLEO \cite{kshort_paper} and 
ALEPH \cite{aleph-97167,aleph_pre} as well as this analysis 
suggest $K_1(1270)$  dominance. An analysis of the $K^-\pi^+$ and
$\pi^-\pi^+$ substructure in $\tau^-\to K^-\pi^+\pi^-\nu_\tau$
allowed ALEPH\cite{tau98aleph} to determine
$f_{1270} = 0.41\pm 0.19\pm 0.15$, based on the known
branching fractions of the $K_1$ resonances to $K^*\pi$ and $K\rho$.
Recent measurements therefore favor $K_1$(1270) dominance in 
 $\tau^-\to K^-\pi^+\pi^-\nu_\tau$.

Calculating the amount of $\rho$ in $K\pi\pi$ from the fit parameters we 
find $f_\rho=0.48\pm0.14$, close to the 
measurement by ALEPH \cite{tau98aleph} of
$f_\rho=0.35\pm 0.11$.
This number also agrees with the measurement by ALEPH of another 
related decay channel,
$\tau^-\to K^0_S\pi^-\pi^0\nu_\tau$ where the component 
$\overline{K}^0\rho^-$ in the intermediate state is found to be
$(64\pm 9\pm 10)\%$, approximately twice that of $K^-\rho^0$
as expected by isospin
symmetries \cite{aleph-97167}. 

\subsection{$K_a-K_b$ mixing}

   From our result for the ratio of $\tau\to K_1\nu_\tau$ decay
amplitudes, information about the mixing of the $K_a$ and
$K_b$  eigenstates can be derived. The mixing between
$K_a$ and $K_b$ is traditionally parameterized
in the following way \cite{suzuki}:
\begin{eqnarray} \nonumber
 K_1(1400) = & K_a \cos{\theta_K} - K_b \sin{\theta_K} \\ \nonumber 
 K_1(1270) = & K_a \sin{\theta_K} + K_b \cos{\theta_K} 
\end{eqnarray}

In the case of exact $SU(3)_f$ symmetry,
the second-class current $\tau\to K_b\nu_\tau$ is forbidden
and only $K_a$ is produced. However, due to the difference
between the masses of the up and strange quarks we may expect
 symmetry breaking effects of  order 
$|\delta| = (m_s-m_u)/\sqrt{2}(m_s + m_u)\approx 0.18$.
Then, instead of pure $K_a$ a 
linear combination $|K_a\rangle - \delta|K_b\rangle$ is produced
and the ratio of decay rates of the $K_1$ resonances can be 
written as \cite{suzuki}:
 
\begin{equation}
\frac{{\cal B}(\tau\to K_1(1270)\nu)}
{{\cal B} (\tau\to K_1(1400)\nu)} = \left|
     \frac{\sin{\theta_K} - \delta\cos{\theta_K}}
     {\cos{\theta_K} + \delta\sin{\theta_K}}
               \right|^2 \times \Phi^2
\label{eq:ratpred}
\end{equation}
In this expression, $\Phi$ is the ratio of appropriate kinematical
and phase space terms and is calculated by numerical integration.
 With the parameters measured in this analysis the ratio of branching 
fractions (Eq. \ref{eq:ratpred}) is written as
\begin{equation}
  \frac{{\cal B}(\tau\to K_1(1270)\nu_\tau)}
           {{\cal B}(\tau\to K_1(1400)\nu_\tau)} = 
              \frac{A^2R_A + C^2R_C}{D^2 R_D}
\label{eq:ratfit}
\end{equation}
From Eqs. (\ref{eq:ratpred})-(\ref{eq:ratfit}), solutions for $\theta_K$
can easily be found: 
\begin{eqnarray} \nonumber
\text{(a) } \theta_K & = (69\pm 16\pm 19)^\circ \text{ for $\delta=0.18$}, \\
\nonumber
\text{(b) } \theta_K & = (49\pm 16\pm 19)^\circ \text{ for $\delta=-0.18$}.
\end{eqnarray}
There is a second pair of solutions that has opposite sign and
the same magnitude.

One can also calculate $\theta_K$ using
the current experimental information on the masses and branching
fractions of $K_1(1270)$ and $K_1(1400)$, independent of their
production in $\tau$-decays. There are two possible
solutions,
$\theta_K\approx 33^{\circ}$ and $\theta_K\approx 57^{\circ}$
\cite{suzuki}. 
   Our result has the same two-fold ambiguity and is consistent
with this calculation.



\section{Summary and Conclusions}

     In this analysis we have measured the relative fractions and
parameters of the $K_1$ resonances in 
$\tau^-\to K^-\pi^+\pi^-\nu_\tau$ decays.
 These measurements are made within the framework of the model
 described in Sec. \ref{sec:our-model}.
Briefly, we assume that $K_1$(1270) and $K_1$(1400) saturate the
$K^-\pi^+\pi^-$ spectrum, and consider only the interference inherent in
the Breit-Wigner mass distributions in calculating the relative
$\tau^-\to K_1(1270)\nu_\tau$ and 
$\tau^-\to K_1(1400)\nu_\tau$ branching fractions.
Our parameterization
 of the axial vector form-factors is different from \cite{mirkes}
 in two respects -- our form-factors
are motivated by isospin relations, and we
assume the Wess-Zumino anomaly to be negligible, 
as described in Sec. \ref{sec:our-model}. 
We find
$f_{1270}=0.66\pm 0.19 \pm 0.13$ and
$f_\rho=0.48\pm 0.14 \pm 0.10$, with 
$f_{1270}$ and $f_\rho$ defined 
as the $K_1(1270)$ and $K\rho$ fractions in $\tau\to K^-\pi^+\pi^-\nu_\tau$.

 These measurements agree well with the recent results from CLEO
 and ALEPH (see Sec \ref{sec:analysisa}).
Our data slightly favor $K_1(1270)$ dominance in production of
the $K^-\pi^+\pi^-$ final state.
The widths that we extract for the $K_1$ resonances are considerably
larger than previously tabulated values\cite{PDG98}.
We also calculate the $K_a-K_b$ mixing angle, finding $\theta_K$ to
be consistent with theoretical expectations.


\acknowledgements

We gratefully acknowledge the effort of the CESR staff in providing us with
excellent luminosity and running conditions.
J.R. Patterson and I.P.J. Shipsey thank the NYI program of the NSF, 
M. Selen thanks the PFF program of the NSF, 
M. Selen and H. Yamamoto thank the OJI program of DOE, 
J.R. Patterson, K. Honscheid, M. Selen and V. Sharma 
thank the A.P. Sloan Foundation, 
M. Selen and V. Sharma thank Research Corporation, 
S. von Dombrowski thanks the Swiss National Science Foundation, 
and H. Schwarthoff thanks the Alexander von Humboldt Stiftung for support.  
This work was supported by the National Science Foundation, the
U.S. Department of Energy, and the Natural Sciences and Engineering Research 
Council of Canada.

\end{document}